# How People Manage Knowledge in their "Second Brains"

## A Case Study with Industry Researchers Using Obsidian

Juliana Jansen Ferreira[1][0000-0003-2867-3568], Vinícius Segura[1][0000-0002-7265-2417], Joana Gabriela Souza[2][0000-0003-1760-1868], and Joao Henrique Gallas Brasil[1]

[1] IBM Research Brazil
[2] Department of Software Engineering, PUC Minas, Belo Horizonte, MG, Brazil
```
{jjansen, vboas} @br.ibm.com, joana.souza@dcc.ufmg.br,
                      jhgb@ibm.com
```

**Abstract.** People face overwhelming information during work activities, necessitating effective organization and management strategies. Even in personal lives, individuals must keep, annotate, organize, and retrieve knowledge from daily routines. The collection of records for future reference is known as a personal knowledge base. Note-taking applications are valuable tools for building and maintaining these bases, often called a "second brain." This paper presents a case study on how people build and explore personal knowledge bases for various purposes. We selected the note-taking tool Obsidian and researchers from a Brazilian lab for an in-depth investigation. Our investigation reveals interesting findings about how researchers build and explore their personal knowledge bases. A key finding is that participants' knowledge retrieval strategy influences how they build and maintain their content. We suggest potential features for an AI system to support this process.

## 1 Introduction

In today's world, people face constant information overload in their work and personal lives. This flow of data creates challenges in capturing and retrieving what is relevant. The volume and variety of information sources complicate identifying what is truly important. Additionally, the nature of information hinders retention and organization for future reference. Thus, the challenge extends to creating efficient systems and strategies for effective retrieval and use.[1,2]

We investigate the records people maintain as their personal knowledge base or "second brain."[3] Research on these bases often draws from Vannevar Bush's MEMEX concept, proposed in his seminal 1945 essay "As We May Think" [4] This hypothetical device was designed to store and retrieve extensive personal information, enabling users to form associative trails between documents and ideas.

Managing a personal knowledge base involves more than organizing knowledge; it extends human cognitive abilities. According to Clark and Chalmers, internal and external entities influence cognition. [5] Using a tool to maintain a personal knowledge base impacts outcomes: "*We shape our tools, and thereafter our tools shape us*." This occurs on cognitive and psychological levels, enhancing or degrading mental capabilities. Furthermore, engaging with a persistent collection of external knowledge can be beneficial by rediscovering, revisiting ideas, and connecting them through associative trails. [3,6,7] Throughout this paper, **personal knowledge base (PKB)** refers to records maintained for future reference, while ***electronic personal knowledge base (ePKB)*** refers to electronic tools for expressing, capturing, and retrieving PKB.

The definition of knowledge has been widely debated. .[8,9] Considering Rowley's definitions [8], we define *knowledge as the combination of data, information, and expert tacit expertise, organized and processed to convey understanding, experience, and learning, resulting in a valuable asset for decision making.*

Note-taking apps, often used as an ePKB, offer intuitive interfaces for efficient note creation, categorization, and organization. Advanced search functions enable easy retrieval of notes, ensuring valuable insights are always accessible. Many apps support multimedia integration, embedding images, audio, and video to enrich notes. Top apps include Notion[1], Microsoft OneNote[2], and Obsidian[3]. These apps empower users to build and maintain comprehensive PKBs.

A "second brain" metaphor describes the ePKB, highlighting its role in enhancing cognition and memory, particularly for note-taking apps. In the Obsidian forum[4], searching for "second brain" reveals plugins, guidelines, and tips related to these features. By capturing and connecting knowledge, note-taking apps serve as an auxiliary brain, enabling users to expand upon their thoughts and ideas. This metaphor emphasizes the potential of note-taking apps to boost productivity, creativity, and lifelong learning as extensions of cognitive abilities.

This paper explores how individuals build and maintain their "second brain", their personal knowledge base (PKB), using a note-taking tool. These apps empower users to build and maintain comprehensive PKBs. We identify users' strategies, practices, and behaviors for organizing, storing, and retrieving information, highlighting knowledge construction and management in digital environments. This investigation contributes to a research project to support the scientific discovery process through a framework that facilitates knowledge consumption and production, incorporating Artificial Intelligence (AI) components to assist discovery tasks. [10,11] We consider PKB content from domain experts [15] as a vital resource for decision-making in the discovery process, serving as proxies for their mental models [17], which helps organize, interpret, and apply information in specific domains. [12] Thus, our framework must account for PKBs in the discovery process.

---

[1]  https://www.notion.so
[2]  https://www.onenote.com
[3]  https://obsidian.md/
[4]  https://forum.obsidian.md/

Our case study comprised interviews with researchers about their use of note-taking apps, especially Obsidian, including the tasks they performed and how they executed them. Our findings reveal how researchers build and use personal knowledge bases (PKB) with a digital note-taking tool (ePKB). One interesting finding is that their planned knowledge retrieval strategies during PKB exploration significantly influence the construction and maintenance of their PKB content. As a result, we provide a comprehensive list of potential features to enhance this process, strengthening our framework's support for effective knowledge management.

This paper is divided into five sections. First, we review the related work on personal knowledge bases and note-taking tools. Next, we explain our case study, with our research methodology. In the subsequent section, we discuss our findings and present a list of potential features for our framework. Finally, in the last section, we present our final remarks, the limitations of our research, and plans for future work.

## 2    Case Study

This work is driven by a long-term research project to conceptualize and develop the XPTO framework. XPTO offers a systematic approach to the discovery process and manages knowledge created throughout its entire lifecycle, from hypothesis generation and testing to conclusion. Additionally, XPTO captures and organizes knowledge acquired during the discovery process for future reuse, thereby expediting future instances of the process.[10,11]

We conducted a case study [13,14] using Obsidian as a note-taking app, with seven Computer Science researchers from a Brazilian research lab who are also Obsidian users as participants. We identified an opportunity for this case study when we noticed that researchers used Obsidian in different and exciting ways. We aimed to learn from their experiences and insights about the tool and its use in daily tasks. As part of the interview, we proposed observation tasks when participants used their Obsidian personal vault to illustrate their discourse throughout the interview.

The authors ensured their research complied with ethics and data privacy guidelines for Research Involving Human Participants. [18] Necessary measures were taken to maintain high ethical standards, including obtaining informed consent, using anonymization to protect identities, and establishing data privacy handling procedures.

This research methodology had four steps[5]:

1. **Define Interview Protocol and Script** - We build a consent form for the study protocol. We manage the consent form and its answers using an online survey software[6]. We designed an open-ended script for interviews aligned with our study goal. The complete consent form and interview script are available here.

2. **Select Participants** - We sent an invitation message to a group of Computer Science researchers we knew were also Obsidian users. The recruitment of participants was voluntary and restricted by the case we wanted to investigate (researchers who were users of Obsidian).

---

[5]   See details here: https://osf.io/a58sr/?view_only=008d50f5d9c24c7c9d4b8ad2ea2d1428
[6]   https://www.alchemer.com/

3. **Conduct Interviews with Observation Tasks** - We conducted one-hour interviews with each participant about their usage of Obsidian in Portuguese. We used open-ended questions to encourage participants to share their experiences and insights about Obsidian usage. At least two research team members participated in each interview, one acting as the lead interviewer and the other as the note-taker. The interviews were conducted using a video system.

**Analyze Study Data** - We performed a content analysis of the data collected during the interviews. [19] After the interviews, we generated video and audio transcriptions using Word for Windows at Microsoft 365[7]. We reviewed and adjusted all the transcriptions before starting the data analysis. To present the findings in this paper, we translated the quotes from Portuguese to English. We aimed to keep the quote translation as accurate as possible to what the participants said. One researcher performed the user study data analysis, but we had discussion sessions about all the interviews with the researchers authoring this paper. The discussions helped us focus on our bigger goal with the case study: understanding how researchers register and retrieve knowledge to build their PKBs.

## 3      Findings

We conducted interviews with the researchers, combined with observation tasks using participants' Obsidian vaults[8]. We present our findings into two categories: ***Creating and Organizing Notes features and Knowledge Retrieval Strategies***. The findings are supported by annotated images from Obsidian's screen and participant interview quotes.

### 3.1     Creating and Organizing Notes features

During the interviews, the creation of notes and how they are organized were recurrent discussion topics. The participants reported the need to keep the context of the knowledge created. When the participant creates a folder or a tag, he communicates that he has defined a context of interest, and the note-creation process should be in that same context. Obsidian is considered a note-taking tool, creating a note is central system functionality, and we will focus on the strategies that the participants can use to structure notes better to recover this information afterward.

We reviewed the Obsidian app to check some noteworthy features that somehow impacted the user interaction related to creating and organizing notes, considering what we heard in interviews, to illustrate our participants' inputs. The user clicks on the folder, "Daily notes" (**Fig. 1**-1), and clicks to create a new note (**Fig. 1**-2).

---

[7] https://insider.microsoft365.com/en-us/blog/transcribe-comes-to-word-for-windows
[8] https://help.obsidian.md/Getting+started/Create+a+vault

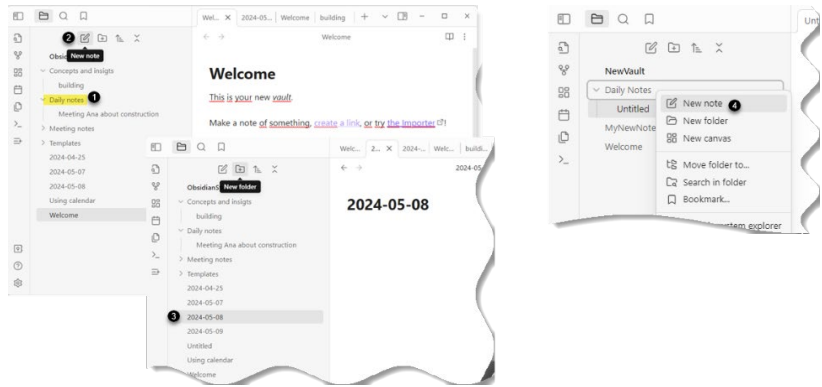

**Fig. 1.** New note created outside the desired folder (1-2-3) + Note created in the folder (4).

This new note is created in the root directory (**Fig. 1**-3), not in the specific folder he wants. If the user selects the folder, he wants to add a note and clicks on the mouse's right button, a "New note" option will appear. This new note will be in the selected folder (**Fig. 1**-4).

Participants reported on the tag feature as a way of organizing notes and relating notes with the same category (e.g., a concept) or subject. This strategy involves adding another layer of meaning and another retrieval system for the notes. Obsidian users can use tags as another way of retrieving and relating their notes, thus making connections that would be more difficult to recover and connect using only the organization based on folders. In Obsidian, a user can use a template created using a markdown to summarize the articles he read for work. He identifies a concept he had seen before, but could not remember exactly what it meant. He searches and creates a new note (**Fig. 2**-1), adds the concept idea in his own words (**Fig. 2**-2), and adds the tags "#concept", "#signs" and "#SemEng" (**Fig. 2**-3), considering that the concept he saved was a sign in the topic of Semiotic Engineering[9]. He returns to the document and puts the "#signs" tag in the header of his template, which appears as a link (**Fig. 2**-4). By clicking on the tag, he can see the note with the concept and open it if he wants. In this case, "sign" (**Fig. 2**-5).

We illustrate below some participants` discourse about how the organization in folders and the use of tags contribute to their PKB construction process:

*I write things down and when I need to make links between them... I actually try to end up keeping everything in folders. This also makes it easier for me to find my notes, right? (P6)*

*(...) having a good tag mechanism, you can build a folder scheme up to dynamic folders... it has a psychological effect on me, I think it's the closest thing to checking an item in a paper list ... is moving something to another folder. There! It is done! (P4)*

---

[9] https://en.wikipedia.org/wiki/Semiotic_engineering

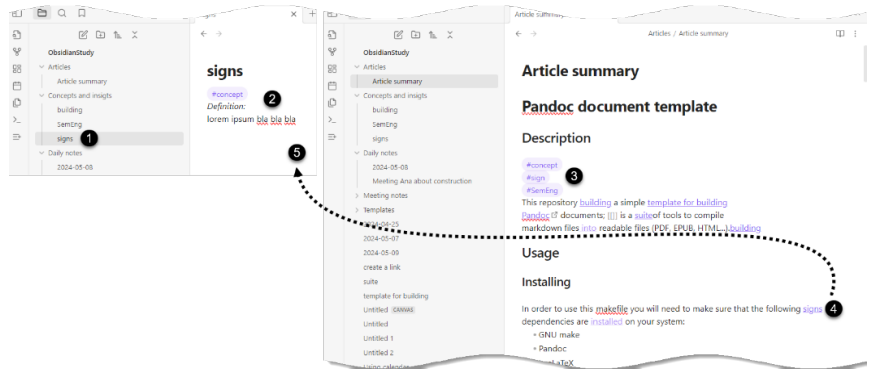

**Fig. 2.** Creating tags and associating them with notes is an organizational strategy.

*(...) I have my tags, they are here (participant shows list of tags in their vault). This here is evidence or this is a conclusion or a definition... (P7)*

*If I think I'm going to forget what that means, or what I wanted to say (...) I put a link there to Slack (messaging tool), which is where there was some discussion of something relevant. (P2)*

*(...) every time I create a note, it goes here to the inbox (folder called ``Inbox''). This inbox, at some point, I process it, let's say once a week.(...) In this experiment, the one that is open, so at some point, I will process it and store it in the appropriate places... (P5)*

### 3.2   Knowledge Retrieval Strategies

We observed that the strategy participants use to create and organize their knowledge in Obsidian is strongly related to their intended retrieval strategy for that knowledge. In this section, we present our findings related to knowledge retrieval strategies.

At Obsidian, we highlighted four retrieval strategies: by searching through the search field (**Fig. 3**-2), by clicking on the tag -- which returns the existing tags (**Fig. 3**-1), within the notes by clicking on tags that appear in the text, or on a link within the notes (**Fig. 3**-3). Considering searching through the search field, it is always visible on the interface, and by clicking on it, the user can click on the tag option that appears in a drop-down menu and select one of the existing tags to retrieve the notes, or type in a word (**Fig. 3**-1). The system will search for the notes in which that word or expression appears. This search strategy demonstrates the tool's concern for offering the user options to quickly find information in a note at the click of a button.

The last two retrieval strategies identified occur within a note. They are more related to the moment when the user is reading or analyzing a note, since they need to consult the information through a link in the note or a tag from the need to understand some extra information or semantic load that can be adapted to the current note from the links and tags (**Fig. 3**-3) that can work as pointers to other related information, explanations, insights, or semantic connections that this extra information can bring to the user.

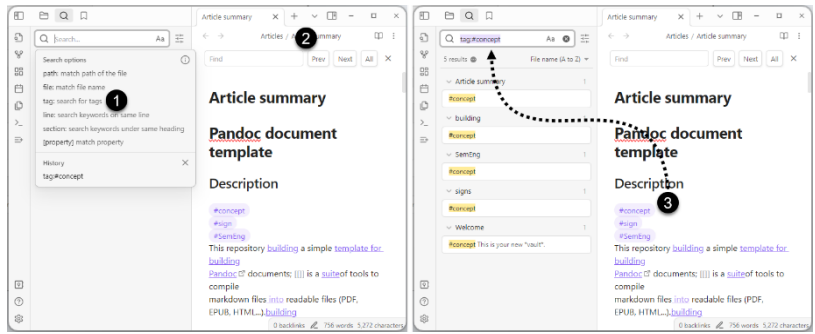

**Fig. 3.** Ways to retrieve knowledge at Obsidian.

In our user study, the knowledge retrieval strategies were discussed, and we illustrate below some users' discourse about how they retrieve previously created knowledge while using Obsidian as their ePKB:

*... I use search... and I use the tags for future me. I try to be nice to him... (P5)*

*(...) you build information, but how did you concatenate it to be useful to you in the future. (P5)*

*(..) if I don't find it, I look for it... I start a search. But normally it's all here. The main things I need to keep are here in the folder structure. (P6)*

*A combination of find and sequential search (..) there are some indexes on these folders, right? So, generally, the indexing of the folders when I had one vault it was okay. Now, with 3 vaults, it's confusing... (P6)*

*It's the search (...) so, if I want ``attention'', I come here ``attention'' (typing into the search box) and then there will normally be one like this (note with ``attention'' concept definitions). Normally, I look for concepts that are true, right? (P1)*

## 4    Discussion

This work is part of the research and development project of a framework to support the creation of discovery applications, enabling knowledge consumption and production. This framework has AI components that aim to support several tasks of the discovery process.[10This case study pointed out some potential features supporting the process. Although AI features were not the focus or discussion topics, we used the insights of this work as references for design ideas and discussions about how AI components can be used in the case.

Obsidian provides the necessary resources of an ePKB for organizing knowledge (like folder tree, tagging, mind and concept maps, hypertext, and flashcards) and retrieving knowledge (search, graphical representation, and spaced repetition).[1] Since retrieval tasks significantly influence the creation and organization strategies, we believe our framework should support the whole process from knowledge creation and organization to knowledge retrieval and use. Considering that, our framework design discussion should consider the following features:

- **Provide a basic knowledge organization structure**—participants reported wanting to create their own organization structure, but starting from a "blank page" might be overwhelming for some users.
- **Allow changes in the knowledge organization structure anytime** - participants provided much evidence about the dynamic knowledge organization process. Therefore, a tool to support that process must support this necessity.
- **Connect creation/organization strategy with retrieval strategy from the start** - provide examples or templates of organization strategies associated with retrieval strategies that may help users relate to a starting template and adjust as necessary.

Participants reported that the main tasks are related to note management, including creating, editing, and recovering notes. These tasks align with the MEMEX idea of registering knowledge and being able to retrieve it in a PKB [4,16]  The relationship between mental models and people's expertise is fundamental to creating, organizing, and retrieving knowledge.[12] In our user study with researchers, we learned how they connect different knowledge sources and how they somehow externalize their tacit knowledge, their mental model about the domain. Moreover, their PKB reflects their discovery process in their expertise domain:

> *(...) place to keep all my thoughts and notes, organized or not (...) I write, I register, then I think about organizing or discarding them  (P3)*

> *I would say that in my case... there are 3 moments, right? The moment I'm writing it on the fly, like in the middle of an experiment, for example. There's a moment when I'm formatting my notes, right? So they can be accessed later. And there is a moment when I have information retrieval. (P5)*

## 5      Final Remarks, Limitations, and Future Work

One important point we need to address regarding this work is the specific profile of our user study participants and the tool we chose for the case study. While our participants perform discovery-related tasks, we recognize that they represent a specific group of Obsidian users (Computer Science researchers).

In future work, we plan to execute a complete and detailed inspection of Obsidian and other note-taking tools to triangulate with these case study findings. In this inspection, we also plan to investigate Obsidian AI resources.[20] In addition, we plan to design and execute studies with expert users from different domains to explore how they build their personal knowledge base to support various tasks related to discovery processes.